\documentclass[english]{article}
\usepackage[T1]{fontenc}
\usepackage{cite}
\usepackage[latin9]{inputenc}
\usepackage{amsmath}
\usepackage{amssymb}
\usepackage{esint}
\makeatletter
\usepackage{color}
\usepackage{epsfig}

\usepackage{graphicx}% Include figure files
\usepackage{dcolumn}% Align table columns on decimal point
\usepackage{bm}% bold math

\usepackage{babel}
\newcommand{\e}{\mbox{e}}

\begin{document}
 
\begin{titlepage}
\thispagestyle{empty}

\bigskip

\begin{center}
\noindent{\Large \textbf
{Cosmologies of Multiple Spherical Brane-universe Model}}\\

\vspace{0,5cm}

\noindent{I. C. Jardim${}^{a}$\footnote{e-mail: jardim@fisica.ufc.br}, R. R. Landim ${}^{a}$\footnote{e-mail: renan@fisica.ufc.br}, G. Alencar ${}^{a}$\footnote{e-mail: geova@fisica.ufc.br} and R. N. Costa Filho${}^{a}$\footnote{e-mail: rai@fisica.ufc.br}}

\vspace{0,5cm}
 
 {\it ${}^a$Departamento de F\'{\i}sica, Universidade Federal do Cear\'{a}-
Caixa Postal 6030, Campus do Pici, 60455-760, Fortaleza, Cear\'{a}, Brazil. 
 }

\end{center}

\vspace{0.3cm}

\begin{abstract}

The Friedmann-Robertson-Walker metric with spherical topology is calculated as an effective metric at the brane in a  multiple branes in $D-$dimensional spacetime scenario. In this model the radius of the brane is the cosmological scale factor, and its evolution is calculated from the viewpoint of the observers on their respective branes. The cosmology of the brane-universe is analyzed in cases where the anisotropic pressure has a cosmological constant like state equation.  The pressure values needed to distinguish the oscillating, expanding or collapsing solutions are found, and the minimum value of a brane-universe mass to prevent a collapse into a black hole is also calculated. Finally the equation of motion is solved numerically in order to illustrate the different cosmological scenarios and to validate the analytical results. 
\end{abstract}
\end{titlepage}

\section{Introduction}
The Friedmann-Robertson-Walker metric (FRW) describes a homogeneous isotropic universe that can be modeled as a perfect fluid. The dynamics of this fluid is given by a cosmological scale factor, which determines the spatial volume of the universe, and depends on the amount of matter in the universe \cite{weinberg:cosmology}. This model became the standard cosmological model due to the observations of Hubble in 1929 \cite{Hubble:1929ig} showing an expanding universe in opposition to predictions of the Einstein's static universe. This observation led to the conclusion that the universe was hotter and denser in the past with an infinite density at the Big Bang. Although such initial singularity causes difficulties to a rational description of the universe, that model gained prominence in the 70's, due to the Hawking and Penrose's singularity theorems \cite{Hawking:1969sw}. However, the latest
observations indicate that the universe is expanding at an increasing rate \cite{Knop:2003iy,Riess:2004nr} what requires some kind of matter which gravitate repulsively like a cosmological constant. This kind of matter violates the conditions of the singularity theorems, making its results inapplicable to the nowadays observed universe. Since then, models without cosmological singularity returned to the scene as possible candidates for the description of the universe like for example models of oscillating
universe \cite{Maeda:2010ke, Hirano:2010sj, Xiong:2008ic, Lessner:2006nq}.

Brane-universe models emerged to solve theoretical problems, like the unification of fundamental interactions and the hierarchy problem \cite{Kaluza:1921tu, Klein:1926tv, ArkaniHamed:1998rs, Antoniadis:1998ig, Randall:1999vf, Randall:1999ee}.  These models describe the universe as a flat brane, and because of this it can not be used to solve cosmological problems, like the accelerated expansion and the initial singularity. Thus, other geometries are necessary but only two are compatible with the FRW metric: the spherical and hyperbolic ones. For easily incorporating the observed symmetries of the universe, like the galaxy isotropic runaway, the existence of a preferred frame, and a cosmic Gaussian time, the spherical brane model has received great attention. The phenomenology of these models has been studied in past decade \cite{Gogberashvili:1998iu,Boyarsky:2004bu} and for more recent studies see \cite{Akama:2011wi,Akama:2012pf,daSilva:2012yd}. These models has been proved compatible with the observational 
data \cite{Tonry:2003zg,Luminet:2003dx,Overduin:1998pn}, as well as, by introduction of different cosmological constant in the bulk, with the observed dynamics, modeling the dark energy \cite{Gogberashvili:2005wy}. 

In this piece of work we continue a recent study where a multiple anisotropic concentric spherical branes model was constructed \cite{Jardim:2011gg}. Here, a dimensional reduction is made to obtain the effective metric measured by observers on the brane and the evolution of the cosmological scale factor in the description of these observers. This reduction give us the cosmology of the brane-universes, in such a way as to be compared with the current data. A more detailed study of the cosmology is made for cases where the anisotropic pressure is of cosmological constant type. This becomes important since this model naturally tends to expand, in agreement with the current observations. Studying the null velocity points we can found the necessary conditions ( but not sufficient) to models of oscillating or eternal expanding universes. In the oscillating scenario we analyze the conditions able to prevent the collapse of the universe into a black hole. To reinforce the study of null velocity points the equation 
of motion is solved numerically. At this point we will discuss the energy condition in the oscillating scenario.

This work is organized as follows: Sec. II gives a short review of results obtained in multiple spherical brane model. In particular the metric, the equation of motion and the possible state equation for anisotropic pressure are shown since they are needed for the sequence of this work. In Sec. III the effective metric on the brane is calculated through a dimensional reduction, and the cosmology of the brane to observers on them is also obtained. In Sec. IV  the possible models of a universe with state equation for anisotropic pressure like a cosmological constant is discussed, and the conditions for the existence of oscillating, collapsing or eternally expanding universes are studied. In Sec. V the equation of motion is solved numerically to illustrate the behavior of the cosmological scale factor
in different scenarios, as well as to confirm the results obtained in Sec. IV. Also in this section we discuss the weak energy condition in oscillating scenario. The last section shows the conclusion and possible perspectives of this work.

\section{The spherical brane model}
Following the multiple spherical brane model developed in ref. \cite{Jardim:2011gg} the Einstein's equation is solved in a $D$ dimensional bulk with spacetime dependent cosmological constant. The matter field has $n$ spherical and anisotropic $(D-2)$-branes, and the obtained invariant line element is     
\begin{equation}\label{dsD}
d\tau^{2} = -A(r,t)dt^{2} +B(r,t)dr^{2} +r^{2}d\Omega^{2}_{D-2},
\end{equation}
where $\Omega_{D-2}$ is the angular line element in $D$ dimensions, formed by $D-2$ angular variables. The metric elements are
\begin{equation}
 B^{-1}(r,t) =  1- \frac{2G_{D}M(r,t)}{r^{D-3}} -r^{2}\lambda(r,t) \label{B} 
\end{equation}
\begin{eqnarray}
A(r,t) &=& B^{-1}(r,t)\prod_{i=0}^{n}\e^{-\pi_{i}\theta(R_{i} -r)}\times \nonumber
\\ && \left[1+\left(\frac{B_{i-1}(R_{i})}{B_{i}(R_{i})}-1\right)\theta(R_{i}-r)\right],\label{A}
\end{eqnarray}
where we defined

\begin{equation}
M(r,t) \equiv  \sum_{i=0}^{n}\frac{R_{i}^{D-2}}{2G_{D}}\left[\frac{2\kappa_{D}\rho_{i}}{(D-2)} -\Delta\lambda_{i}R_{i}\right]\theta(r-R_{i}),\label{Mrt}
\end{equation}

\begin{equation}
\lambda(r,t) \equiv \sum_{i=0}^{n}\lambda_{i}[\theta(r -R_{i}) - \theta(r -R_{i+1})],
\end{equation}

\begin{eqnarray}
B_{j}^{-1}(r) &=& 1- \frac{1}{r^{D-3}}\sum_{i=0}^{j}\left[\frac{2\kappa_{D}}{D-2}\rho_{i}R_{i}^{D-2} -\Delta\lambda_{i}R^{D-1}_{i}\right] -\nonumber
\\&& -\lambda_{j}r^{2},\label{Bj}
\end{eqnarray}
and
\begin{equation}
\pi_{i} \equiv \frac{2\kappa_{D}}{D-2}R_{i}B_{i}(R_{i})P_{i},
\end{equation}
where $P_{i}$ is the anisotropic pressure, 2$\Delta\Lambda_{i} = (D-1)(D-2)\Delta\lambda_{i}$ is the difference between the internal and external cosmological constant of the $i$-th brane, and for completeness $M_{0} = R_{0} =0$. Since the branes are dynamic, the following equation of motion for the $i$-th brane is obtained from the conservation of energy-momentum tensor:
\begin{eqnarray}
\rho_{i}\frac{dU_{i}}{dt_{i}}&=& \frac{\Delta\Lambda_{i}}{\kappa_{D}}\left(1- U_{i}^{2}\right) -\frac{D-2}{R_{i}}\left[P_{i} -\rho_{i}\left(\gamma_{i}
+U_{i}^{2}\right)\right] - \nonumber
\\&&-B_{i}(R_{i})\left[P_{i} +\rho_{i} -2\rho_{i}U_{i}^{2}\right]\times \nonumber
\\&&\times\left[(D-3)\frac{G_{D}M(R_{i})}{R_{i}^{D-2}} -R_{i}\lambda_{i}\right],\label{evolution}
\end{eqnarray}
where
$
dt_{i} = \sqrt{A_{i}(R_{i})/B_{i}(R_{i})}dt $,
$
U_{i} \equiv \frac{d R_{i}}{dt_{i}} $ and $\gamma_{i}$ is the ratio between the pressure and energy density of matter on the brane. The achievement of the above equation of motion gives the following two state equations that remove divergences
\begin{equation}
P_{i} = -\rho_{i} \;\;\;\;\;\mbox{or}\;\;\;\;\;  P_{i} = \rho_{i}U_{i}^{2}. \label{cond}
\end{equation}
With the above anisotropic pressures the brane can have any value of matter on it. This freedom allows us to keep the cosmological eras, in same way the standard cosmological model. 

\section{Effective metric and cosmology on the brane}
In order to calculate the effective metric and its evolution measured by observers on a specific brane a reduction is performed to describe the brane-universes cosmology. To do this we will start with the
$D$-dimensional metric (\ref{dsD}), and using the relationship $r=R_{i}(t)$, we can make the dimensional reduction on the brane to obtain the effective metric
\begin{equation}
 ds_{i}^{2} = -\left[A_{i}(R_{i}) -B_{i}(R_{i})V_{i}^{2}\right]dt^{2} +R_{i}(t)^{2}d\Omega_{D-2}^{2}.
\end{equation}
To simplify the above expression, we use the following comoving time and radial coordinate on the brane
\begin{eqnarray}
d\tau_{i} &=& \sqrt{A_{i}(R_{i}) -B_{i}(R_{i})V_{i}^{2}}dt,
\\ \bar{r} &=& \sin\theta_{0}.
\end{eqnarray}
In this coordinates, the reduced metric takes the form
\begin{equation}
 ds_{i}^{2} = -d\tau^{2}_{i} +R_{i}(\tau_{i})^{2}\left[\frac{d\bar{r}^{2}}{1-\bar{r}} +\bar{r}^{2}d\Omega_{D-3}^{2}\right].
\end{equation}
This is the Friedmann-Robertson-Walker metric for a universe with $(D-2)$ spatial dimensions with spherical topology, where the brane radius is the cosmological scale factor. To obtain the scale factor evolution, $R_{i}(\tau_{i})$, we need to calculate the evolution law (\ref{evolution}) in terms of the new variable  $\tau_{i}$. Using the composed derivation rules we obtain
\begin{eqnarray}
U_{i} &=& \sqrt{\frac{B_{i}(R_{i})}{1 +B_{i}(R_{i})W_{i}^{2}}}W_{i},
\\ \frac{dU_{i}}{dt_{i}} &=& \frac{B_{i}(R_{i})}{(1 +B_{i}(R_{i})W_{i}^{2})^{2}}\left[\frac{dW_{i}}{d\tau} +\frac{B_{i}(R_{i})'}{2B_{i}(R_{i})}W_{i}^{2} \right],
\end{eqnarray}
where $W_{i} = dR_{i}/d\tau_{i}$ and the prime indicates the derivation with respect to the argument. In terms of comoving time the equation of motion (\ref{evolution}) takes the form
\begin{equation}\label{evoi}
\rho_{i}\frac{dW_{i}}{d\tau_{i}} = b_{0}(R_{i}) +b_{2}(R_{i})w_{i}^{2} +b_{4}(R_{i})w_{i}^{4},
\end{equation}
where, to simplify, we defined $w_{i} \equiv \sqrt{B_{i}(R_{i})}W_{i}$. The above coefficients for the $P_{i} = -\rho_{i}$ case are
\begin{eqnarray}
b_{0}(R_{i}) &=& \frac{\Delta\Lambda_{i}}{\kappa_{D}B_{i}(R_{i})} +(D-2)\rho_{i}\frac{1 +\gamma_{i}}{B_{i}(R_{i})R_{i}}, \nonumber
\\b_{2}(R_{i}) &=& \frac{\Delta\Lambda_{i}}{B_{i}(R_{i})\kappa_{D}} +\frac{(D-2)}{R_{i}}\rho_{i}\left[3 +\frac{2\gamma_{i}}{B_{i}(R_{i})}\right]- \nonumber
\\&&-3(D-1)\rho_{i}\left[\frac{G_{D}M(R_{i})}{R_{i}^{D-2}} +\lambda_{i}R_{i}\right],\label{b01}
\\b_{4}(R_{i}) &=& \frac{(D-2)}{R_{i}}\rho_{i}\left[2 +\frac{\gamma_{i}}{B_{i}(R_{i})}\right]- \nonumber
\\&& -2(D-1)\rho_{i}\left[\frac{G_{D}M(R_{i})}{R_{i}^{D-2}} +\lambda_{i}R_{i}\right], \nonumber
\end{eqnarray}
and, for $P_{i} = \rho_{i}U_{i}^2$, the coefficients are
\begin{eqnarray}
b_{0}(R_{i}) &=& \frac{\Delta\Lambda_{i}}{\kappa_{D}B_{i}(R_{i})} +(D-2)\frac{\rho_{i}\gamma_{i}}{B_{i}(R_{i})R_{i}}-\nonumber
\\&& -\rho_{i}\left[(D-3)\frac{G_{D}M(R_{i})}{R_{i}^{D-2}} -R_{i}\lambda_{i}\right] ,\nonumber
\\b_{2}(R_{i}) &=& \frac{\Delta\Lambda_{i}}{B_{i}(R_{i})\kappa_{D}} +\frac{2(D-2)\gamma_{i}\rho_{i}}{B_{i}(R_{i})R_{i}}, \label{b02}
\\b_{4}(R_{i}) &=& \frac{(D-2)\rho_{i}\gamma_{i}}{B_{i}(R_{i})R_{i}}. \nonumber
\end{eqnarray}
Equation (\ref{evoi}) gives us the brane-universe cosmology as felt by observers on the brane, and as one can see, from velocity independent coefficient, the cosmologies are strongly influenced by the choice of radial pressure. In the first case (Eqs. (\ref{b01})), the brane-universe tends to expand due to the negative pressure ( like a  cosmological constant), while the in the second case (Eqs. (\ref{b02})) it tends to collapse.

\section{Cosmology for $P_{i} = -\rho_{i}$}
In this section we study the model with a cosmological constant like state equation. We choose the  $P_{i} = -\rho_{i}$ case because it has a state equation that simplifies the analytical calculations.  Furthermore, this state equation is consistent with the energy-momentum tensor used in the model, and provides a natural explanation for the condition expressed in Eq. (\ref{cond}). There are two different cosmological scenarios depending on the signal of acceleration when the velocity of the $i$-th brane vanishes. That happens for $w_{i}=0$, and the equation (\ref{evoi}) give us the acceleration 
\begin{eqnarray}
\left.\rho_{i}\frac{dW_{i}}{dt_{i}}\right|_{W_{i} =0}&=& \frac{\Delta\Lambda_{i}}{\kappa_{D}B_{i}(R_{i})} +(D-2)\rho_{i}\frac{1 +\gamma_{i}}{B_{i}(R_{i})R_{i}}.\label{return}
\end{eqnarray}
Considering only the region where the energy density, $\rho_{i}$, is positive, the acceleration will be positive if
\begin{eqnarray}
0 &<& \frac{\Delta\Lambda_{i}}{\kappa_{D}} +\frac{D-2}{R_{i}}\left(1 +\gamma_{i}\right)\rho_{i} \nonumber
\\&\propto&  \Delta\lambda_{i}R_{i}^{D-1}\left[(D-1) +(D-2)\left(1 +\gamma_{i}\right)\right] +\nonumber
\\&&+2(D-2)\left(1 +\gamma_{i}\right)G_{D}M_{i}.
\end{eqnarray}
Assuming that $\gamma_{i}\geq -1$, what is physically consistent since we don't know anything that  violates this condition, the acceleration in the rest
points will be positive if
\begin{equation}
-\frac{2(D-2)\left(1 +\gamma_{i}\right)G_{D}M_{i}}{(D-1) +(D-2)\left(1 +\gamma_{i}\right)} < \Delta\lambda_{i}R_{i}^{D-1}.
\end{equation}
For the $\Delta\lambda_{i} >0$ case, the solution is trivial ($R_{i}>0$), and this universe model will expand indefinitely, which makes it consistent with the observations. For the $\Delta\lambda_{i} <0$ case we obtain the non-trivial solution 
\begin{equation}
 R_{i} < \left[\frac{2G_{D}M_{i}}{|\Delta\lambda_{i}|}\left(1 +\frac{D-1}{(D-2)\left(1 +\gamma_{i}\right)}\right)^{-1}\right]^{1/D-1}.
\end{equation}
The above result shows that, in rest points, the acceleration will be positive if the brane radius is smaller than the above quantity, and vice-versa. The obtained
result defines a critical radius
\begin{equation}\label{RC}
 R_{i}^{c} =\left(\frac{2G_{D}M_{i}}{|\Delta\lambda_{i}|\sigma_{i}}\right)^{1/(D-1)},
\end{equation}
where
\begin{equation}
\sigma_{i} \equiv 1 +\frac{D-1}{(D-2)\left(1 +\gamma_{i}\right)}.
\end{equation}
The adopted procedure don't give us the return points, but show the existence of a rest point with negative acceleration beyond the critical radius and other
with positive acceleration before it. This condition is required for an oscillating universe model. However, the equation of motion (\ref{evoi}) is valid just beyond
the Kottler horizon. Therefore, in order to have an oscillating universe is necessary that the critical radius be bigger than the horizon radius,
\begin{equation}
R^{k}_{i} <  \left(\frac{2G_{D}M_{i}}{|\Delta\lambda_{i}|\sigma_{i}}\right)^{1/(D-1)}.
\end{equation} 
Since the horizon is defined by the pole of the metric (\ref{B}), given by equation
\begin{equation}\label{horizonte}
 -\lambda_{i} \left(R^{k}_{i}\right)^{D-1} +  \left(R^{k}_{i}\right)^{D-3} -2G_{D}\sum_{j=0}^{i}M_{j} =0,
\end{equation}
we can find the mass that makes the critical radius outer the horizon, which is
\begin{equation}
\left(\frac{2G_{D}M_{i}}{|\Delta\lambda_{i}|\sigma_{i}}\right)^{-2/(D-1)} - \left(|\Delta\lambda_{i}|\sigma_{i}
+\lambda_{i}\right) > |\Delta\lambda_{i}|\sigma_{i}\frac{M_{i-1}^{int}}{M_{i}},
\end{equation}
where
\begin{equation}
 M_{i-1}^{int} \equiv \sum_{j=0}^{i-1}M_{j}.
\end{equation}
The above expression has analytical solution just for the most inner brane, where $M_{i-1}^{int} = 0$. To this brane the critical radius will be bigger than the horizon
only if
\begin{eqnarray}\label{MC}
M_{1} &<& \frac{|\Delta\lambda_{1}|}{2G_{D}}\sigma_{1} \left(|\Delta\lambda_{1}|\sigma_{1} +\lambda_{1}\right)^{-(D-1)/2}.
\end{eqnarray}
This is the maximum mass where the most inner brane can have in order to not collapse. If the mass is bigger than this value the evolution law forces that it is contracting and
pass through the horizon creating a black hole. If the mass is less than its maximum value the brane can oscillate as in bouncing models. 

\section{Numerical Results}
In this section the equation of motion (Eq. (\ref{evoi})) is solved numerically. The parameters values  used in the calculations were chosen in a way to show the general behavior expected according for different scenarios shown in last section. We should point that the numerical results
are more complete than the analytical ones. In this case we have found just a necessary condition to distinguish between the different scenarios, but here we find 
the complete solution once the initial conditions are given.

\begin{figure}
\centering
 \includegraphics[scale=0.4]{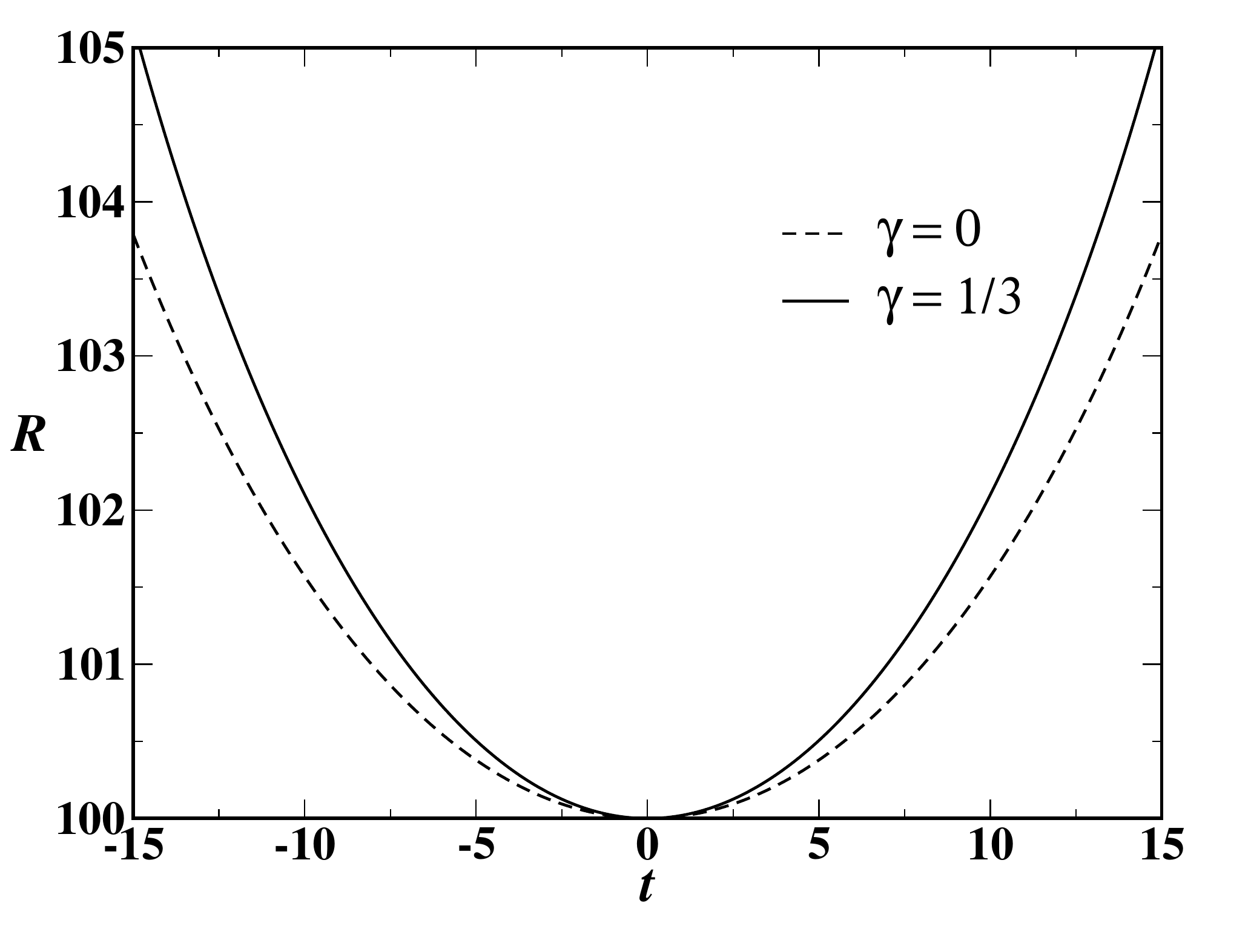}
  \caption{Numerical solution for $D=5$, $G_{5}M = 2$ and $\kappa_{5} = 1$. The dashed line is the solution for dust matter on the brane while the solid line is for a photon gas. Here we considered an expansive scenario with $\lambda_{0} = \lambda_{1} = 0$  and   $R_{i}(0) = 100$ and $W_{i}(0) =0$ as initial conditions. }\label{fig1}
 \end{figure}

\begin{figure}
\centering
 \includegraphics[scale=0.4]{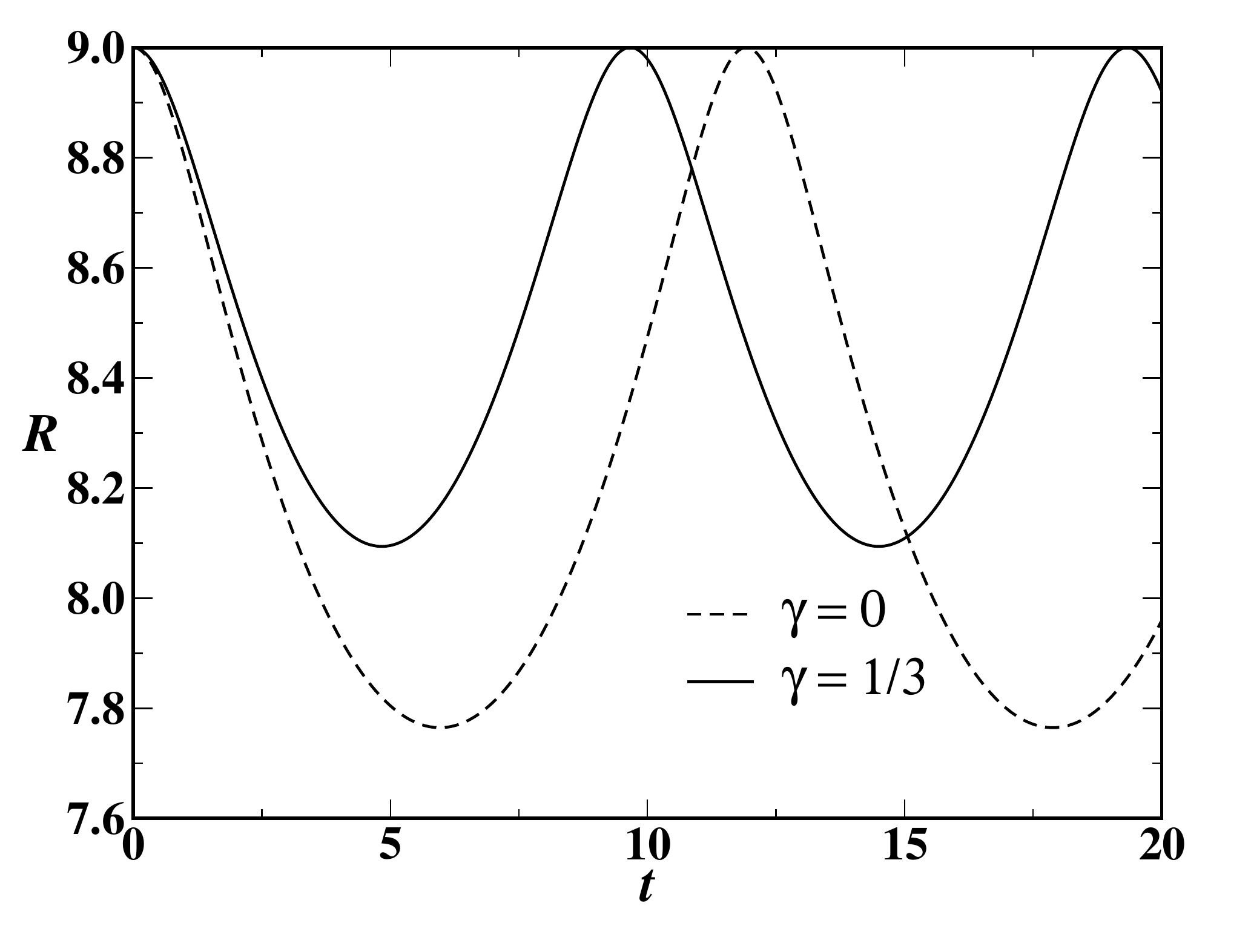}
  \caption{Numerical solution for $D=5$, $G_{5}M = 2$ and $\kappa_{5} = 1$. The dashed lines are the solutions for dust matter on the brane while the solid line is for a photon gas. we simulate an oscillating scenario,  for this we use $ \lambda_{1} = 0 $ and $ \lambda_{0} = 4 \times 10^{-4} $ 
 whit $R_{i}(0) = 9.0$ and $W_{i}(0) = 0$ as initial conditions.} \label{fig2}
\end{figure}

Fig. \ref{fig1}  shows the result for a scenario without cosmological constant, i.e., an expanding universe model. The curve shows the behavior of the radius $R$ against time. From the figure, one can see an expansionary phase preceded by a collapsing stage indicating a model with bouncing.  For the used parameters the universe expands rapidly with time. In Fig. \ref{fig2} we consider a scenario of an oscillating universe, in accordance with the parameters obtained previously. The result shows that the oscillation amplitude must be limited. This happens because when considering a negative difference of cosmological constants (needed for an oscillating scenario) the energy density 
on the brane becomes non-positive defined. This can be seen from the definition (\ref{Mrt})

\begin{figure}
\centering
 \includegraphics[scale=0.4]{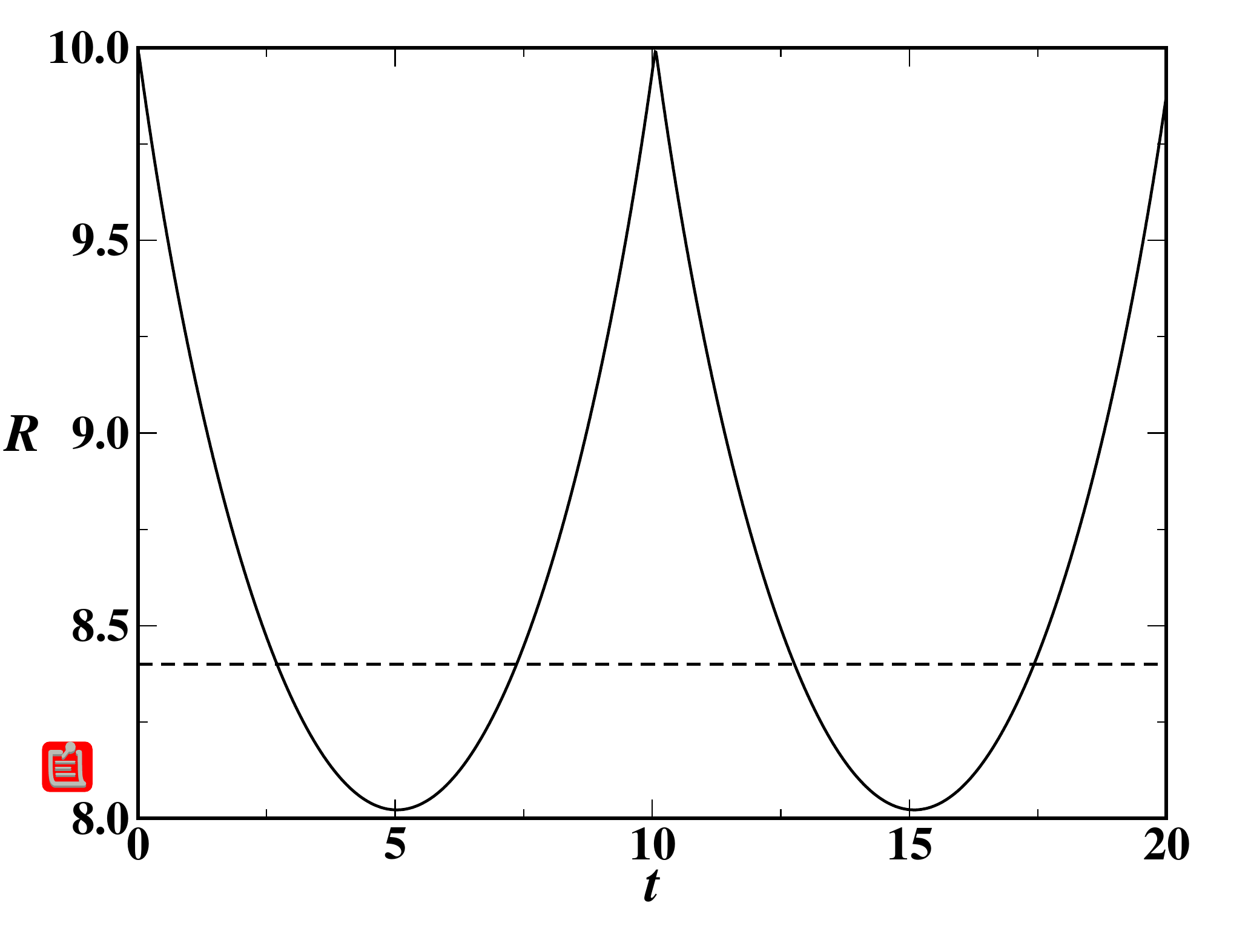} 
  \caption{In above simulation we use the same values of Fig. \ref{fig2} but we fix $R_{i}(0) = 9.99$ and $\gamma = 1/3$. The dashed line 
 indicates the critical radius $R_{i}^{c}$. This illustrate the different behavior for the regions below and above the critical radius.}\label{fig3}
\end{figure}
\begin{equation}
\rho_{i} = \frac{(D-2)}{2\kappa_{D}R_{i}^{D-2}}\left[2G_{D}M_{i} -|\Delta\lambda_{i}|R_{i}^{D-1}\right].
\end{equation}
In order to ensure the weak energy condition \cite{wald:GR} we have to limit the brane radius to
\begin{equation}
R_{i} \leq \left[\frac{2G_{D}M_{i}}{|\Delta\lambda_{i}|}\right]^{1/D-1},  
\end{equation}
so, we can overestimate the oscillation amplitude 
\begin{equation}
 A_{i} \lessapprox 2\left(R_{i}^{\max} - R_{i}^{c}\right) = 2R_{i}^{\max}\left[1- \sigma_{i}^{-1/(D-1)}\right],
\end{equation}
assuming the strong energy condition, $ \gamma_ {i} \leq 1/3 $, the above result leads us, in $ D = 5 $, to
\begin{equation}
 A_{i} \approx  0.32\left(\frac{2G_{5}M_{i}}{|\Delta\lambda_{i}|}\right)^{1/4}.
\end{equation}
The parameters used in the simulation give us the value $A_{i} \lessapprox 3.2$. Fig. \ref{fig2} shows an amplitude of $\approx 0.8$, within the estimated range. The difference occurs because we considered $R_{i}(0) = 9.0$ ( while $\rho_{i} \to 0$ when $R_{i} \to 10.0$) and the behavior for the region $R_{i} >R^{c}_{i}$ is different of the behavior for $R_{i} < R^{c}_{i}$, as we can see in Fig. \ref{fig3} where results for $R_{i}(0) = 9.99$ provides us with an amplitude of $\approx 2.0$, which is still much lower than the overestimated result. These results show the validity of the analytical study of the previous section.

\section{Conclusions and Perspectives}
In this work we compute the effective metric measured by observers on the brane. We obtained, in agreement with the standard cosmological model, the Friedmann-Robertson-Walker
metric with spherical topology. We also obtained the cosmology of these brane-universe, that describes the evolution of the cosmological scale factor. We observed that
the brane cosmology is affected by bulk cosmological constant, by matter on the brane, as well as by the brane state equation. The matter dependence give us the idea
of cosmological eras, which depends on the state equation of this matter.

We also study the cosmology obtained from evolution equation (\ref{evolution}) in the
case where the anisotropic pressure has a cosmological constant like state equation. We show that this kind of brane tends to expand, due the negative pressure, so it is
able to describe the universe without the introduction of cosmological constants in the bulk. In case where the difference of cosmological constants in
the bulk is positive or null we found an eternal expanding universe. An interesting scenario was found using an negative difference between cosmological constants. This has the effect of inhibit the expansion and we obtained an oscillating universe. In this scenario we also obtain the mass limit which a brane could have in order to not collapse into a black hole. Finally, we solve numerically the equation of motion in order to illustrate and reinforce the analytical results. In the oscillating case we estimate the oscillating amplitude 
imposing the strong energy condition in the brane. 

An extension of this work is the study of cosmology generated by the velocity dependent state equation and study the phenomenology of the model. Despite its advantages, we should point that the spherical brane scenario share some problems of plane brane models. One of this problems is related to the localizability of fields on the brane \cite{Alencar:2010vk,Landim:2011ki,Landim:2011ts,Alencar:2012en,Fonseca:2011ep,Fu:2012sa}. This can have important phenomenological implications to the standard model and cosmology \cite{Csaki:2002gy,Csaki:1999mp, Csaki:1999jh}. 
However, in the present manuscript we have supposed that the matter is localized in the brane and we leave the study of trapping of fields in the spherical brane scenario for a future work.

\section*{Acknowledgments}

We acknowledge the financial support provided by Funda\c c\~ao Cearense de Apoio ao Desenvolvimento Cient\'\i fico e Tecnol\'ogico (FUNCAP), the Conselho Nacional de 
Desenvolvimento Cient\'\i fico e Tecnol\'ogico (CNPq) and FUNCAP/CNPq/PRONEX.

\end{document}